\documentclass[12pt, twoside]{article}
\usepackage{a4wide,amssymb,cite}
\usepackage{epsfig}
\usepackage[usenames,dvipsnames]{color}
\usepackage{slashed}

\newcommand{\be}{\begin{equation}}
\newcommand{\ee}{\end{equation}}
\newcommand{\bea}{\begin{eqnarray}}
\newcommand{\eea}{\end{eqnarray}}

\begin{document}

\color{black}
\begin{flushright}
CERN-PH-TH/2012-136\\
\end{flushright}

\vspace{1cm}
\begin{center}
{\huge\bf\color{black} Fermi Gamma Ray Line at 130 GeV  \\[3mm] from Axion-Mediated Dark Matter }\\
\bigskip\color{black}\vspace{1.5cm}{
{\large\bf Hyun Min Lee$^{a}$, Myeonghun Park$^{a}$ and Wan-Il Park$^{b}$}
\vspace{0.5cm}
} \\[7mm]

{\em $(a)$ {CERN, Theory Division, CH--1211 Geneva 23,  Switzerland.}}\\
{\it $(b)$ School of Physics, KIAS, Seoul 130-722, Korea.  }\\
\end{center}
\bigskip
\centerline{\large\bf Abstract}
\begin{quote}\large
We consider a singlet Dirac fermion with Peccei-Quinn(PQ) symmetry as dark matter.
A singlet complex scalar is introduced to mediate between dark matter and the SM through Higgs portal interaction and electroweak PQ anomalies. We show that a resonant annihilation of dark matter with axion mediation can explain the monochromatic photon line of the Fermi LAT data at $130\,{\rm GeV}$ by anomaly interactions while the annihilation cross section with Higgs portal interaction is p-wave suppressed.
We discuss the interplay between the direct detection of the fermion dark matter and the collider search of Higgs-like scalars. We also present a ultra-violet completion of the dark matter model into the NMSSM with PQ symmetry.

\end{quote}

\thispagestyle{empty}

\normalsize

\newpage

\setcounter{page}{1}

\section{Introduction}

%Dark matter
Dark matter is a dominant component of matter density in the universe, occupying about 5 times larger than baryonic matter \cite{dm}. It is known that dark matter(DM) interacts with the Standard Model particles gravitationally, while the property of dark matter such as mass and other interactions has not been known. Weakly Interacting Massive Particles (WIMP) has been a paradigm for cosmology, explaining the dark matter relic density by thermal freezeout with weak-scale mass and weak interactions for dark matter. Thus, it is expected that WIMP is detectable in underground direct detection experiments \cite{xenon}. Recently, from the direct production of dark matter, the LHC has provided a new constraint on dark matter models, in particular, from the limit on spin-dependent cross section of dark matter \cite{lhcdm}.

%Previous Fermi LAT
While DM direct searches have imposed strong limits on the direct detection cross section, indirect searches from the cosmic gamma ray such as the Fermi Large Area Telescope (LAT) \cite{fermilat} has reached the sensitivity to DM annihilation cross section close to $\langle\sigma v\rangle=3\times 10^{-26}\,{\rm cm}^3\,{\rm s}^{-1}$, the one at DM freezeout, in the case of the purely s-wave annihilation \cite{fermilowDMmass}. The gamma ray constraints on the monochromatic photons coming from the DM annihilation have been reported from the Fermi LAT with 11 months and 24 months data \cite{fermi11months, fermi24months}.
A recent tentative analysis of the gamma ray data from the Fermi LAT
\cite{ibarra, weniger} has shown an indication for a gamma ray line at $E_\gamma\approx 130\,{\rm GeV}$ at $4.6\sigma$ local significance\cite{weniger}. 
If the photon excess is explained by dark matter annihilating into a photon pair,
the observations hint at a dark matter mass of $m_X=129.8\pm 2.4^{+7}_{-13}\,{\rm GeV}$ and a partial annihilation cross section of $\langle\sigma v\rangle_{XX\rightarrow \gamma\gamma}=(1.27\pm 0.32^{+0.18}_{-0.28})\times 10^{-27}{\rm cm}^3\, {\rm s}^{-1}$ or 
$\langle\sigma v\rangle_{XX\rightarrow \gamma\gamma}=(2.27\pm 0.57^{+0.32}_{-0.51})\times 10^{-27}{\rm cm}^3\, {\rm s}^{-1}$, depending on the dark matter profile\footnote{See also a different analysis of the Fermi LAT data \cite{raidal}}. 
The results are very interesting, although it needs a further confirmation for excluding the possibility of instrumental errors or unknown astrophysical backgrounds.

%DM model
On the DM model building side, however, it is a nontrivial task to obtain such a large branching fraction into monochromatic photons from the annihilation of dark matter, because the relevant annihilation channels are loop-suppressed \cite{nmssm, inertdoublet, singletscalar, extradm, mambrini, servant, effDM, boxshape, cline, kiyoung, kyae}.  Furthermore, since the light SM degrees of freedom produced from the DM annihilation generates a continuum spectrum of gamma rays, one has to check if the continuum spectrum is consistent with the line spectrum. In this work, we propose a new dark matter model to explain the branching fraction into monochromatic photons as indicated by the Fermi LAT data \cite{weniger} and discuss the gamma ray constraints in the model.

%Model
We consider a fermion dark matter which carries PQ charge and couples to a complex scalar singlet. The complex scalar singlet does not couple to the SM directly. 
Instead, the real part of the complex scalar (CP-even scalar) mediates dark matter interactions to the SM Higgs boson by Higgs portal interaction, so does the imaginary part (the so called axion in our model) to the SM electroweak gauge bosons by PQ anomalies. 
We show that the DM annihilation with axion mediation is a dominant channel in determining the relic density at freezeout by a resonance effect and it produces a monochromatic photon line by the DM annihilation into $\gamma\gamma$ or $Z\gamma$ with sizable branching fractions at present. On the other hand, the DM annihilation with CP-even scalar mediation can contribute to the total cross section at freezeout comparably to the one with axion mediation while becoming p-wave suppressed at present. 
As a consequence, we show that there is a parameter space that explains the observed Fermi gamma ray line at $E_\gamma\approx 130\,{\rm GeV}$ and satisfies all the phenomenological constraints such as gamma ray constraints, DM direct detection and the Higgs-like scalar search at the LHC, etc. 

Although the introduction of extra DM annihilation coming from the CP-even singlet scalar reduces the gamma-ray line to the central values as given in Ref.~\cite{weniger}, extra contribution does not have to be sizable, given the systematic and statistical errors of the DM annihilation cross section reported. In other words, 
the predicated gamma-ray line of our model without CP-even scalar mediation is still consistent with DM annihilation into two photons (at about 2 sigma level with the NFW dark matter profile \cite{weniger}), so is it with DM annihilation into $Z\gamma$ for any dark matter profile.  But, we have included the CP-even singlet sector in our DM discussion as it naturally appears in the PQ completion of  the axion-like scalar.

The paper is organized as follows.
First, we begin with a description of our model with PQ symmetric dark sector.
Then, we present the results of the DM relic density and the production mechanism of gamma ray line and discuss the direct detection constraint on the model.
We proceed to present a ultra-violet(UV) complete model for generating the electroweak anomalies by the electroweak PQ axion in the context of the NMSSM.
Finally, a conclusion is drawn.
There are two appendices providing the details of the decay rates of the CP-even scalars and the axion, and the cross sections for dominant DM annihilation channels with axion and CP-even scalar mediations.

\section{The Model}

We consider a Dirac fermion dark matter $\chi$ that is charged under $U(1)_{PQ}$ global symmetry.
When a complex singlet scalar $S$ couples to the fermion dark matter, it mediates the dark matter interactions to the Higgs boson $H$ by the Higgs portal interaction and the Standard Model electroweak gauge bosons $A^i_\mu$ by anomalies. The SM Higgs doublet does not have a PQ charge because of the Yukawa couplings \cite{kimaxion}.
We define PQ transformations on dark matter and the mediator as $\chi\rightarrow e^{i\gamma_5\alpha}\chi$ and $S\rightarrow e^{-2i\alpha} S$, respectively.

We specify the relevant action for dark matter to be the following,
\be
{\cal L}= i{\bar\chi}\gamma^\mu\partial_\mu\chi+|\partial_\mu S|^2+|D_\mu H|^2-V(H,S)-\frac{1}{4}F^i_{\mu\nu} F^{i\mu\nu}+{\cal L}_{\rm int}
\label{action}
\ee
with
\bea
V(H,S)&=& \lambda_H |H|^4+\lambda_S |S|^4+2\lambda_{HS}|S|^2 |H|^2 +m^2_H |H|^2+m^2_S|S|^2-\Big(\frac{1}{2}m^{\prime 2}_S S^2+{\rm h.c.}\Big), \label{potential} \\
{\cal L}_{\rm int}&=&\lambda_\chi (S {\bar \chi} P_L\chi +S^* {\bar\chi}P_R \chi )+ \sum_{i=1,2,3}\frac{c_i\alpha_i}{ 8\pi  v_s}\, a F^i_{\mu\nu} {\tilde F}^{i\mu\nu}
\eea
where ${\tilde F}_{\mu\nu}\equiv\frac{1}{2}\epsilon_{\mu\nu\rho\sigma}F^{\rho\sigma}$, $v_s\equiv \sqrt{2} \,\langle S\rangle$ is the axion decay constant and we need $|\lambda_\chi|\lesssim {\cal O}(1)$ for a valid effective theory for dark matter with mass $M_\chi\equiv\lambda_\chi v_s/\sqrt{2}$, and the constant parameters $c_i$ depend on the anomalies generated by the axion-like scalar. If dark matter has a fixed mass as hinted by the tentative result from the Fermi LAT data, the dark matter coupling $\lambda_\chi$ is determined by the axion decay constant $v_s$. We take $v_s$ to be larger than the axion mass such that the anomaly loops with heavy fermions having masses of order $v_s$ are well approximated by the dimension-5 interactions between the axion and the SM electroweak gauge bosons.

The interactions of a complex scalar field $S=(s+ia)/\sqrt{2}$ containing $a$ as the imaginary part generates the DM mass and DM-axion coupling in the above action.
When the anomalies with nonzero $c_i$ are generated by the couplings of the complex scalar $S$ to extra heavy fermions with SM charges, the axion mediates the dark matter interactions to the SM gauge bosons only through 
anomalies\footnote {We note, however, that in the SM with two Higgs doublets $H_u$ and $H_d$, the Higgs doublets can have nonzero PQ charges \cite{dineaxion} such that a renormalizable coupling, $SH_u H_d$ or $S^2 H_u H_d$, could be written depending on the PQ charge assignments. In this case, the axion could mix with the pseudo-scalar Higgs, acquiring a tree-level coupling to the SM particles. As will be discussed in a later section, such couplings can be chosen to zero in a PQ supersymmetric NMSSM.}.
We assume that our axion does not couple to colored fermions so it can be called the electroweak axion. Otherwise, the DM annihilation into a gluon pair would be too large to give a sizable branching fraction into photons.
Since the anomaly interactions are model dependent, we treat them to be arbitrary parameters. We postpone the discussion on microscopic models to a later section.

For the PQ mechanism to work for solving the strong CP problem, the QCD anomalies must be generated by the invisible axion that couples to extra heavy quarks \cite{kimaxion}. 
We note that the axion-like scalar $a$ gets a PQ-breaking mass $m'_S$ from a higher dimensional interaction with a PQ breaking scalar $\Phi$ containing the invisible axion with $\langle \Phi\rangle=F_a$: for $V=-\frac{1}{2M^2_P}\Phi^4 S^2+{\rm h.c.}$, we get $m'_S=\frac{F^2_a}{M_P}$. Thus, for $F_a \sim 10^{10}\,{\rm GeV}$, which is within the invisible axion window, $10^9\,{\rm GeV}<F_a<10^{12}\,{\rm GeV}$, the Planck suppressed term generates a weak-scale mass for the axion-like scalar. Furthermore, the invisible axion can constitute the dark matter relic density too, but we assume that it is subdominant.
On the other hand, if the soft PQ-breaking mass violates CP, a mixing between CP-even and -odd scalars could lead to too large branching fraction of the DM annihilation into the SM particles.
In order for the axion to couple to the SM only by anomalies, we assume that the induced PQ breaking mass does not violate CP.

After minimizing the potential in eq.(\ref{potential}), the VEVs of the singlet and the Higgs doublet, $v_s$ and $v$, are determined as \cite{lebedev}
\bea
v^2_s&=&  \frac{\lambda_{HS} m^2_H-\lambda_H (m^2_S-m^{\prime 2}_S)}{\lambda_S\lambda_H-\lambda^2_{HS}},\\
v^2&=&  \frac{\lambda_{HS}(m^2_S-m^{'2}_S)-\lambda_S m^2_H}{\lambda_S\lambda_H-\lambda^2_{HS}}.
\eea
The conditions for a local minimum are $\lambda_{HS}m^2_H-\lambda_H (m^2_S-m^{\prime 2}_S)>0$, $\lambda_{HS}(m^2_S-m^{\prime 2}_S)-\lambda_S m^2_H>0$ and $\lambda_S \lambda_H-\lambda^2_{HS}>0$.
Expanding the scalar fields around the vacuum as $S=(v_s+s+ia)/\sqrt{2}$ and  $H^T=(0,v+h)/\sqrt{2}$ in unitary gauge,
the obtained mass matrix for CP-even scalars can be diagonalized by the field rotation,
\be
s=\cos\theta\, {\tilde s}+\sin\theta\,{\tilde h}, \quad h=-\sin\theta\,{\tilde s}+\cos\theta\,{\tilde h}
\ee
with
\be
\tan\,2\theta=\frac{2\lambda_{HS} v_s v}{\lambda_H v^2-\lambda_S v^2_s},
\ee
and the mass eigenvalues are
\bea
m^2_{1,2}=\lambda_H v^2+ \lambda_S v^2_s\mp \sqrt{(\lambda_S v^2_s-\lambda_H v^2)^2+4\lambda^2_{HS} v^2 v^2_s}. \label{masses}
\eea
Due to the singlet-Higgs mixing\footnote{ Recently, even a small singlet-Higgs mixing can lead to a sizable threshold correction to the Higgs quartic coupling at tree-level, solving the vacuum instability problem of the Higgs mass lighter than $130\,{\rm GeV}$ in the SM \cite{vacuumstability}. }, the real scalar also mediates the dark matter interactions to the SM particles as the Higgs boson does. 
The axion mass is just given by the PQ breaking mass as $m_a=m'_S$.
In our model, we can trade four independent Lagrangian parameters in the CP-even scalar sector to four physical parameters by eqs. (4)-(8): $v_s, m_{1,2}$ and the mixing angle $\theta$.

%%%%%%%%%%%%% Begin OF FIGURE ################
\begin{figure}[t]
\centering%
\includegraphics[width=16.5cm]{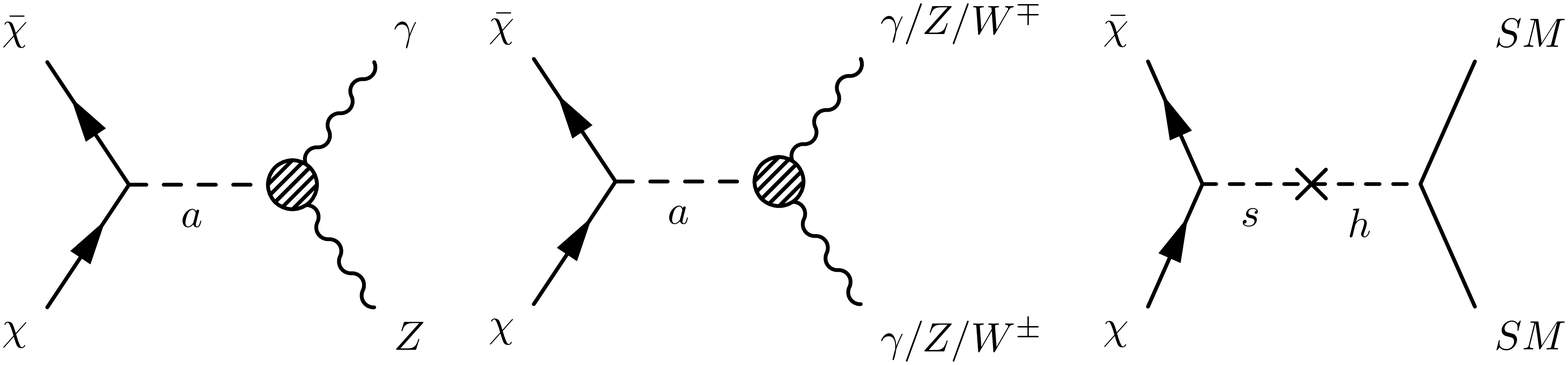}
\caption{Feynman diagrams for dark matter annihilation with axion and CP-even scalar.}\label{fig:diagram}
\end{figure}
%%%%%%%%%%%%% End   OF FIGURE ################

\section{Dark Matter Constraint and Fermi Gamma Ray Line}

In this section, we consider the constraint coming from the DM relic density in our model and  show how a monochromatic photon line is produced by DM annihilation as observed by the Fermi LAT. We also discuss the interplay between DM direct detection and Higgs search at the LHC in the interesting region that explains the Fermi gamma ray line.

\subsection{Dark matter annihilation cross section}

Dark matter annihilates into an SM pair through both the axion and the real scalar partner of the axion.
There are four channels with s-channel axion in our model as shown in Fig.~1: ${\bar\chi}\chi\rightarrow a\rightarrow\gamma\gamma, Z\gamma, ZZ, W^+ W^-$. All the channels are s-wave so their annihilation cross sections are little changed since the freezeout.  The first two channels lead to monochromatic gamma lines from dark matter annihilation. If the branching fraction of the first two channels is sizable and DM annihilation occurs due to the resonance at $m_a\sim 2 M_\chi$, it is possible to explain the tentative gamma ray line observed at Fermi LAT with the cross section required to explain the dark matter relic density. We denote the annihilation cross section coming from axion mediation by $\langle \sigma v\rangle_a$.

%%%%%%%%%%%%% Begin OF FIGURE ################
\begin{figure}[t]
\centering%
\includegraphics[width=7cm]{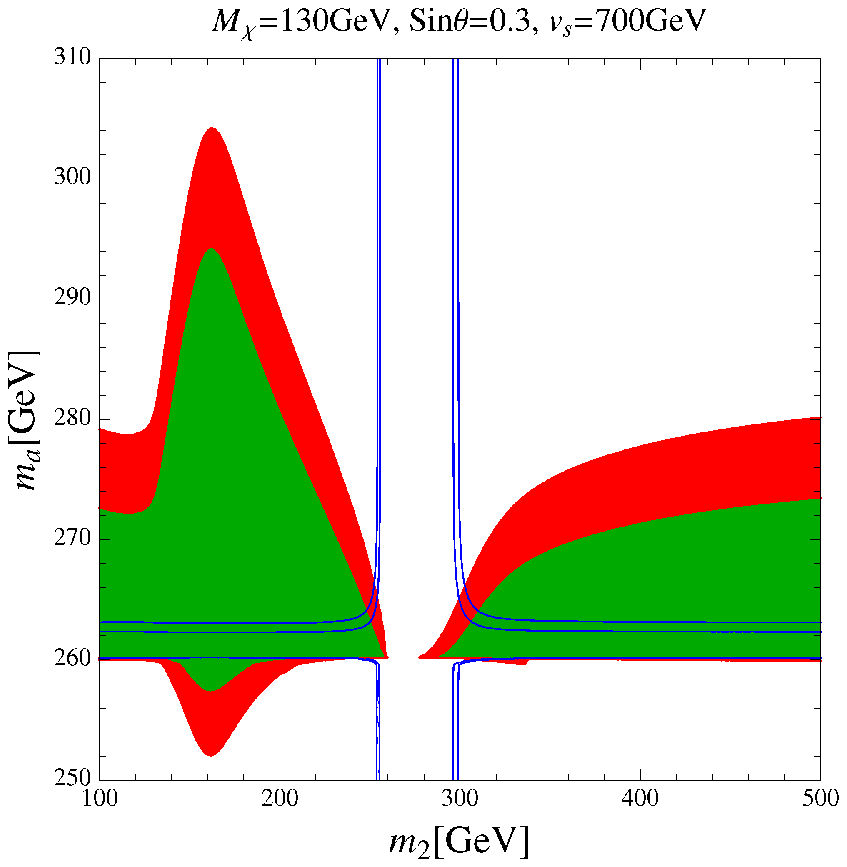}
\includegraphics[width=7cm]{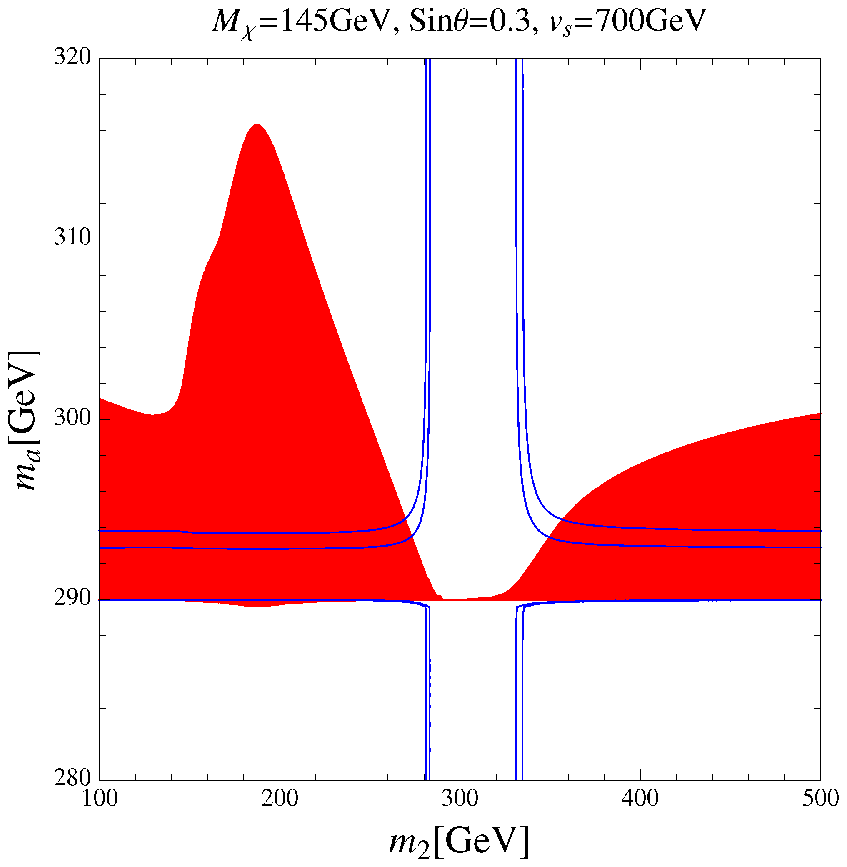}
\caption{Parameter space of $m_a$ vs $m_2$ consistent with WMAP $3\sigma$ band of the relic density (Blue). Left: Parameter space of the present partial annihilation cross section into a photon pair within $4-8\%$ (Red) and larger than $8\%$ (Green) and $c_1=c_2=1$. Right: Parameter space of the present partial annihilation cross section into $Z\gamma$ within $8-13\%$ (Red) and $c_1=0.2, c_2=1$. The mass of the Higgs-like scalar is chosen to $m_1=125\,{\rm GeV}$.}
\label{fig:WMAP}
\end{figure}
%%%%%%%%%%%%% End   OF FIGURE ################

On the other hand, dark matter can also annihilate by the mixing between the CP-even singlet scalar and the Higgs boson. If kinematically allowed, there are four channels with s-channel CP-even scalar annihilating into the SM particles as shown in Fig.~1: ${\bar\chi}\chi\rightarrow s\rightarrow {\bar f}f, ZZ, W^+ W^-, hh$. As shown in appendix B, all the channels are p-wave suppressed. So, even if they can be relevant for generating the correct relic density 
at freezeout\footnote{Dirac fermion dark matter with a CP-even singlet scalar mediation was previously considered\cite{pko}.}, they become suppressed at later time. We denote the annihilation cross section coming from CP-even scalar mediation by $\langle\sigma v\rangle_s$. 
Moreover, singlet self-interaction and/or Higgs portal interaction could lead to extra annihilation channels at tree level: ${\bar\chi}\chi\rightarrow ss, sh, hh, aa, as$ in the interaction basis. All the additional channels except ${\bar\chi}\chi\rightarrow as$ are p-wave suppressed. In particular, the first three channels into CP-even scalars can be comparable to the loop-induced channels with axion mediation at freezeout while they do not affect the branching fraction of the annihilation cross section into monochromatic photons. 
On the other hand, the ${\bar\chi}\chi\rightarrow aa$ channel is kinematically forbidden even at freezeout, close to $m_a= 2M_\chi$ that we are interested in for gamma ray line. The ${\bar\chi\chi\rightarrow as}$ channel is s-wave so the cascade annihilation of the CP-even scalar into the SM particles could produce too many intense secondary photons \cite{fermilowDMmass}. 
So, we forbid the ${\bar\chi\chi\rightarrow as}$ channel kinematically by taking $m_a+m_{1,2}>2 M_\chi$ and search the parameter space that is compatible with the relic density.   
Therefore, the total annihilation cross section is given approximately by the addition of the axion-mediated and scalar-mediated cross sections at freezeout, 
\be
\langle\sigma v\rangle|_{\rm fr}\simeq\langle\sigma v\rangle_a|_{\rm fr}+\langle\sigma v\rangle_s|_{\rm fr}.
\ee

Then, from the velocity times cross section of dark matter annihilation, $\sigma v=a+b v^2$, the dark matter relic density is given by
\begin{equation}
\Omega_{\chi}h^2=\frac{2.09\times 10^8\,{\rm GeV}^{-1}}{M_{Pl}\sqrt{g_{*s}(x_F)}\,\left(a/x_F+3b/x^2_F\right)} \, ,
\end{equation}
where the freeze-out temperature gives $x_F=M_\chi/T_F\approx20$
and $g_{*s}(x_F)$ is the number of the effective relativistic degrees of freedom entering in the entropy density.
The DM relic density measured by WMAP is $\Omega_{\rm DM} h^2=0.1123\pm 0.0035$  at $68\%$ CL \cite{dm}.

In Fig.~\ref{fig:WMAP}, on the $m_a-m_2$ mass plane, for the fixed DM mass to $M_\chi=130\,{\rm GeV}$ or $145\,{\rm GeV}$, we show the parameter space satisfying the WMAP bound on the relic density at $3\sigma$ and the branching fraction of the DM annihilation into a photon pair within $8\%$ or into $Z\gamma$ within $8-13\%$ at present. 
We have used the Feynrules \cite{feynrules} and micrOmegas \cite{micromegas} for the numerical analysis. 
There are four separate lines satisfying the relic density. Each pair of lines parallel to the $m_2$ axis gets close to the axion resonance due to a small decay width of the axion. The necessary tuning for the axion resonance to produce the correct relic density within the WMAP $3\sigma$ band is $\frac{\delta m_a}{m^0_a}\simeq 0.008$ for $m_a=m^0_a+\delta m_a$, that is, $\delta m_a\simeq 2\,{\rm GeV}$ for $m^0_a=2m_\chi=260\,{\rm GeV}$. 
Each pair of lines parallel to the $m_a$ axis are separated wider from each other due to a larger decay width of the CP-even scalars.
We note that there is a parameter space in red in both figures where the gamma ray line at $130\,{\rm GeV}$ can be obtained with the observed intensity as will be discussed in more detail later.

\subsection{Gamma ray line spectra and differential flux of photons}

The spectrum of the $\gamma\gamma$ line is a delta function at $M_\chi$.
On the other hand, the photon spectrum per annihilation with the process ${\bar\chi}\chi\rightarrow Z \gamma$ depends on the mass and width of Z-boson as follows \cite{servant},
\be
\frac{dN^Z_\gamma}{dE_\gamma}=\frac{4M_\chi M_Z \Gamma_Z}{f_1 f_2}
\ee
where $\Gamma_Z$ is the decay width of the Z-boson and 
\bea
f_1&\equiv & \tan^{-1}\Big(\frac{M_Z}{\Gamma_Z}\Big)+\tan^{-1}\Big(\frac{4M^2_\chi-M^2_Z}{M_Z \Gamma_Z}\Big), \\
f_2&\equiv & (4M^2_\chi - 4M_\chi E_\gamma -M^2_Z)^2+\Gamma^2_Z M^2_Z.
\eea

The differential photon flux produced by dark matter annihilation and integrated over the region of angular size $\Delta \Omega$ is computed as \cite{servant}
\be
\frac{d\Phi_\gamma}{dE_\gamma}(E_\gamma)=\frac{1}{4\pi} \frac{r_\odot \rho^2_\odot}{4M^2_\chi} \frac{dN_\gamma}{dE_\gamma} {\bar J}\Delta\Omega
\ee
with
\bea
\frac{dN_\gamma}{dE_\gamma}&=&2\pi \langle \sigma v\rangle_{\gamma\gamma}\,\delta(E_\gamma-M_\chi) +\langle\sigma v\rangle_{Z\gamma}\,\frac{dN^Z_\gamma}{dE_\gamma}, \\
{\bar J}&=& \frac{1}{\Delta\Omega}\int_{\Delta\Omega} J(\psi),\quad J(\psi)=\int_{\rm los}\frac{ds}{r_\odot} \Big(\frac{\rho(r(s,\psi))}{\rho_\odot}\Big)^2.
\eea
Here, we note that $\rho({\vec x})$, $\rho_\odot=0.3\,{\rm GeV/cm}^3$ and $r_\odot=8.5\,{\rm kpc}$ denote the dark matter density at a location $\vec x$ with respect to the GC, its value at the solar system and the distance between the Sun and the GC. And the coordinate $s$ spans along the line of sight, making an angle $\psi$ with the direction of the GC. In order to compare to the spectrum observed by Fermi LAT, we should take into account a $10\%$ energy resolution so we can replace the actual spectrum by a Gaussian distribution function centered at the photon energy. 

There are models for the dark matter density distribution in our galaxy. 
The Navarro Frenk and White (NFW) profile is often used for indirect searches,
\be
\rho_{NFW}(r)=\frac{\rho_s}{\frac{r}{r_s} \Big(1+\frac{r}{r_s}\Big)^2}\,
\ee
On very small scales, the optimal fit to simulated DM halos is provided by the Einasto profile,
\be
\rho(r)=\rho_0\,{\rm exp}\bigg[-\frac{2}{\alpha} \Big(\Big(\frac{r}{R}\Big)^\alpha-1\Big)\bigg]
\ee
where $\alpha=0.17$ and $R=20{\rm kpc}$.
The Einasto profile is shallower than NFW at very small radii.

\subsection{Monochromatic gamma ray from fermion dark matter}

In this subsection, we consider the annihilation of the fermion dark matter with axion mediation and show that the monochromatic photons with sizable branching fraction can be produced.

Suppose that the gamma ray line at $E_\gamma=130\,{\rm GeV}$ hinted by the recent analysis \cite{weniger} is explained by dark matter with $M_\chi=130\,{\rm GeV}$ annihilating into two photons.  Then, there is another peak at $E_\gamma=M_\chi\Big(1-\frac{M^2_Z}{4M^2_\chi}\Big)\simeq 114\,{\rm GeV}$. 
On the other hand, if the observed gamma ray line is explained by the DM annihilation into one photon, we need to choose the dark matter mass to $M_\chi=145\,{\rm GeV}$ and another peak coming from the DM annihilation into two photons is at $E_\gamma=145\,{\rm GeV}$.

%%%%%%%%%%%%% Begin OF FIGURE ################
\begin{figure}[t]
\centering%
\includegraphics[width=7cm]{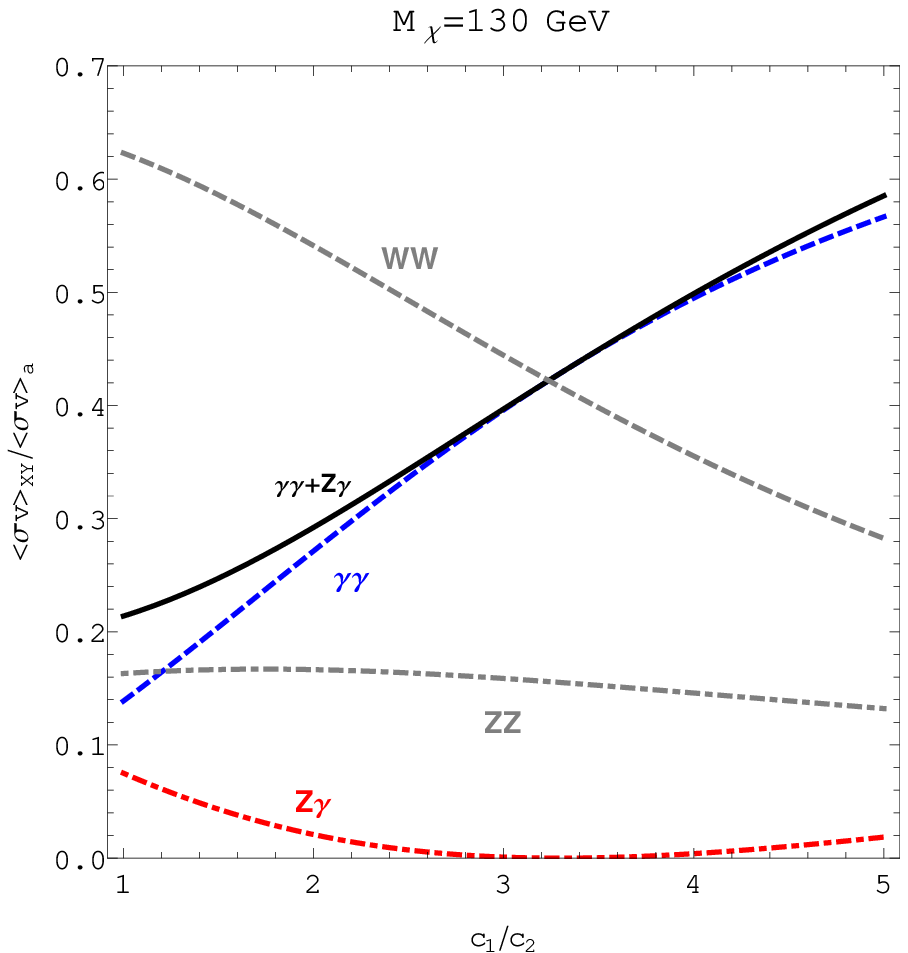}
\includegraphics[width=7cm]{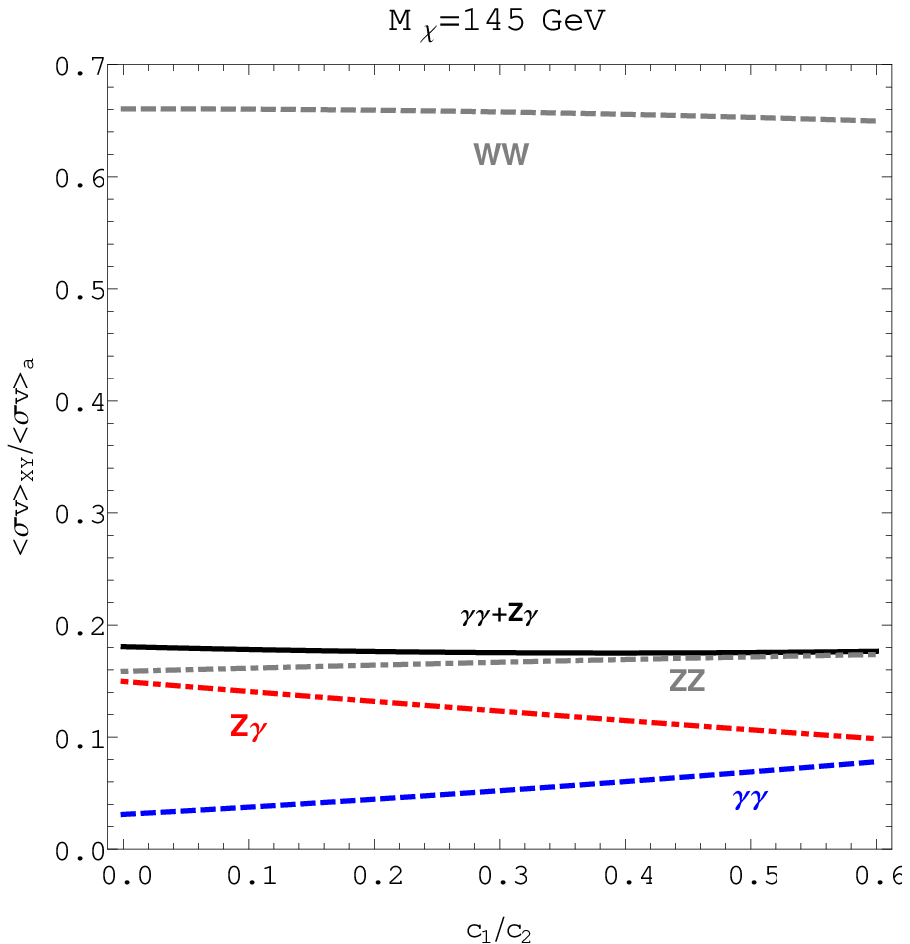}
\caption{Branching fraction of the partial cross sections for DM annihilations with axion mediation vs $c_1/c_2$. Dark matter mass is fixed to give the dominant photon line.}\label{fig:BR}
\end{figure}
%%%%%%%%%%%%% End   OF FIGURE ################

The ratio of the annihilation cross sections into $Z\gamma$ and $\gamma\gamma$ in our model is determined by
\be
r\equiv \frac{\langle\sigma v\rangle_{Z\gamma}}{\langle \sigma v\rangle_{\gamma\gamma}}=\frac{(c_2\alpha_2-c_1\alpha_1)^2\sin^2(2\theta_W)}{2(c_1\alpha_1\cos^2\theta_W+c_2\alpha_2\sin^2\theta_W)^2}\,\Big(1-\frac{M^2_Z}{4M^2_\chi}\Big)^3.
\ee
For $c_1=c_2$, which is the case where two Higgsinos generate electroweak anomalies in the NMSSM as will be discussed in the later section, the above ratio becomes $r\simeq 0.54$ for $M_\chi=130\,{\rm GeV}$.  Then, the intensity of the one-photon line coming from $Z\gamma$ is then $0.27$ times that of the 2-photon line.
Although the one-photon peak may be resolved from the two-photon peak separated by $16\,{\rm GeV}$, the suppressed one-photon line may not be significant due the background while the 2-photon line can explain the observed Fermi gamma ray line\footnote{ It has been shown that two photon lines are shown to fit the gamma-ray data equally well \cite{twophotons}. In particular, if $\langle \sigma v\rangle_{Z\gamma}/(2\langle \sigma v\rangle_{\gamma\gamma})\lesssim 1$, the region around dark matter mass $130\,{\rm GeV}$ remains the best fit.}.
On the other hand, for $c_1=0.2 c_2$, the cross section ratio becomes $r\simeq 2.9$ for $M_\chi=145\,{\rm GeV}$. Then, the intensity of the two-photon line coming from $\gamma\gamma$ is $0.69$ times that of the 1-photon line.  But, in this case, the 1-photon line fits the Fermi gamma-ray line worse than the case with 2-photon line dominance \cite{two photons}.
The branching fraction for the annihilation cross sections is depicted in Fig.~\ref{fig:BR} for $M_\chi=130\,{\rm GeV}$ and $M_\chi=145 \,{\rm GeV}$, depending on $c_1/c_2$.

%%%%%%%%%%%%% Begin OF FIGURE ################
\begin{figure}[t]
\centering%
\includegraphics[width=7cm]{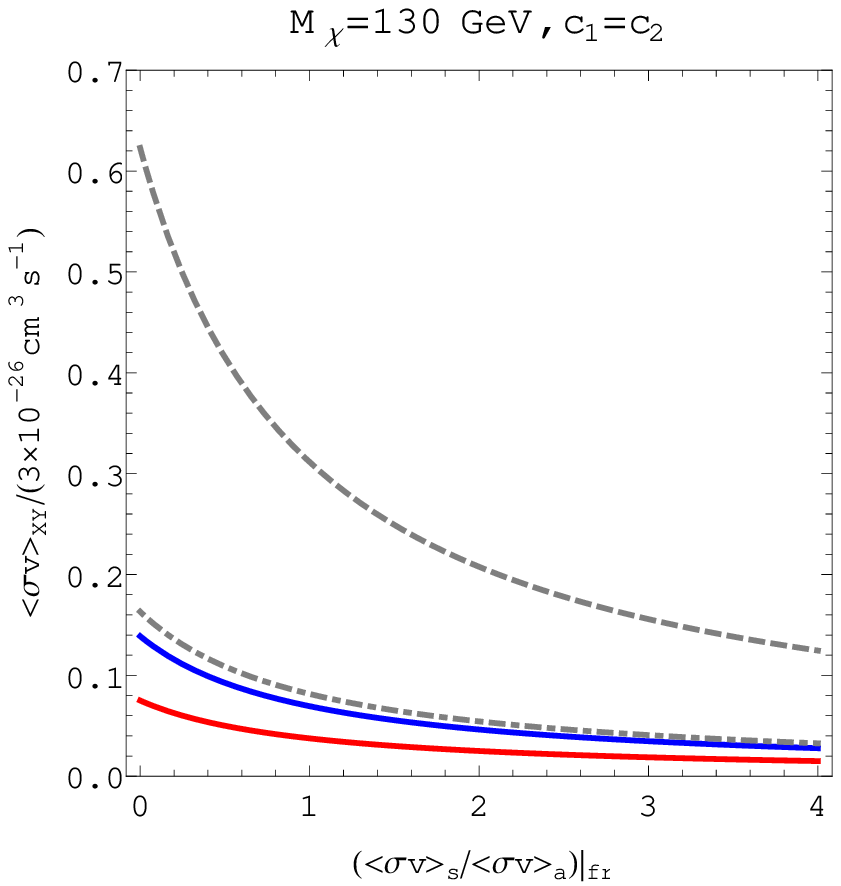}
\includegraphics[width=7cm]{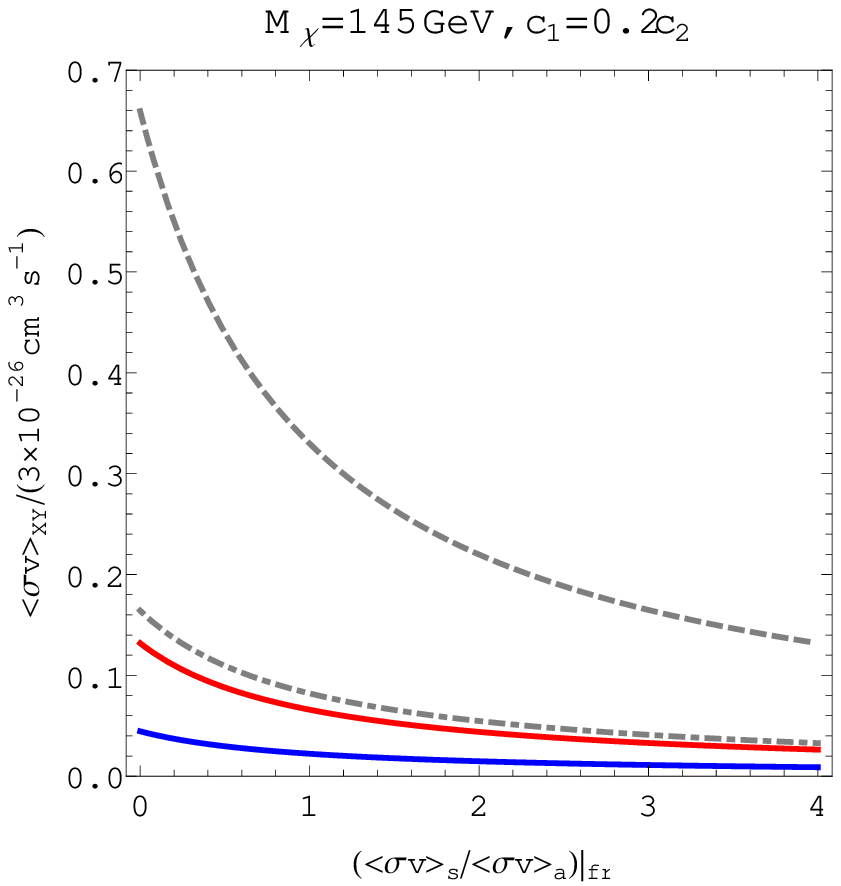}
\caption{Partial annihilation cross sections vs ratio of CP-even scalar to axion-mediated cross sections at freezeout. Dark matter mass is fixed to give the dominant photon line. Left: $WW, ZZ, \gamma\gamma, Z\gamma$ from top to bottom. Right: $WW, ZZ,  Z\gamma, \gamma\gamma$ from top to bottom. }\label{fig:sigmav}
\end{figure}
%%%%%%%%%%%%% End   OF FIGURE ################

In order to explain the observed Fermi gamma ray line by dark matter annihilation,
one needs the cross section of dark matter annihilating into a pair of monochromatic photons to be $\langle\sigma v\rangle_{\gamma\gamma}=(1.27\pm 0.32^{+0.18}_{-0.28})\times 10^{-27}{\rm cm}^3 {\rm s}^{-1}$ for the Einasto profile and
$\langle\sigma v\rangle_{\gamma\gamma}=(2.27\pm 0.57^{+0.32}_{-0.51})\times 10^{-27}{\rm cm}^3 {\rm s}^{-1}$ for the NFW profile, that is, ${\rm Br}({\bar\chi}\chi\rightarrow \gamma\gamma)\simeq 4-8\%$ for thermal dark matter \cite{weniger}.
In our model, the DM annihilation with axion mediation is s-wave so it alone would have produced a more intense gamma ray line than observed because the branching fraction into photons is rather large as shown in Fig.~\ref{fig:BR}.  
But, the DM annihilation channels with CP-even scalar mediation can give a sizable contribution to the total cross section at freezeout while they do not affect the branching fraction into photons at present because of a p-wave suppression. 
Thus, as $\langle\sigma v\rangle_s=0$ at zero temperature, the total annihilation cross section at present is less than the one at freezeout so we can reduce the partial annihilation cross section into photons at present. We can write the partial cross sections at present in terms of the total annihilation cross section at freezeout,
\be
\langle\sigma v\rangle_{XY}=\frac{{\rm Br}({\bar\chi}\chi\rightarrow XY)}{1+\frac{\langle\sigma v\rangle_s}{\langle\sigma v\rangle_a}\Big|_{\rm fr}}\cdot \langle \sigma v\rangle_{\rm fr}\,, \quad {\rm Br}({\bar\chi}\chi\rightarrow XY)\equiv\frac{\langle\sigma v\rangle_{XY}}{\langle\sigma v\rangle_a}.
\ee
We plot the present partial annihilation cross sections with respect to the ratio of scalar-mediated to axion-mediated cross sections at freezeout in Fig.~\ref{fig:sigmav}.
For $c_1=c_2$, the DM annihilation into two photons is dominant to explain the Fermi LAT peak and we get ${\rm Br}({\bar\chi}\chi\rightarrow \gamma\gamma)\simeq 0.14$.
In this case, we need the annihilation cross section with CP-even scalar mediation at freezeout to be $\langle\sigma v\rangle_s/\langle\sigma v\rangle_a= 0.8-2.3$ for $\langle\sigma v\rangle_{\gamma\gamma}/(10^{-27}{\rm cm}^3 {\rm s}^{-1})=1.27-2.27 $.
On the other hand, for $c_1=0.2 c_2$, the DM annihilation into one photon can explain the Fermi LAT peak and we get ${\rm Br}({\bar\chi}\chi\rightarrow Z\gamma)\simeq 0.13$. So,  we need a smaller annihilation cross section with CP-even scalar mediation as compared to the case with two photons: $\langle\sigma v\rangle_s/\langle\sigma v\rangle_a= 0-0.54$ for $\langle\sigma v\rangle_{Z\gamma}/(10^{-27}{\rm cm}^3 {\rm s}^{-1})=2.54-3.9$.

We note that the partial cross section into $WW$ is the largest, being about 5 times larger than the one for the dominant photon channel. But, it satisfies the current limit of about $10^{-25}{\rm cm}^3 \,{\rm s}^{-1}$, coming from the gamma ray emission of dwarf spheroidal galaxies observed 
by the Fermi LAT \cite{fermilowDMmass}. Furthermore, PAMELA has measured the spectrum of cosmic anti-proton flux below $180\,{\rm GeV}$, which is consistent with the background \cite{pamela}. In our model, the anti-proton flux coming from the hadronic decays of $WW, ZZ$ with the annihilation cross sections of about $60\%$ and $20\%$, respectively, is well below the measured one by one or two orders of magnitude \cite{cmantiparticle}.

\section{Direct detection and LHC Higgs search}

As discussed in the previous section, although the DM annihilation with axion mediation gives  a sizable branching fraction of the annihilation cross section into monochromatic photon(s), the level of the gamma ray line excess of the Fermi LAT data requires a sizable contribution of the CP-even scalar mediation to the total annihilation cross section at freezeout. This is achieved by a sizable mixing between the CP-even scalar singlet and the Higgs boson. Then, the same mixing parameter determines the direct detection cross section of dark matter.

%%%%%%%%%%%%% Begin OF FIGURE ################
\begin{figure}[t]
\centering%
\includegraphics[width=7cm]{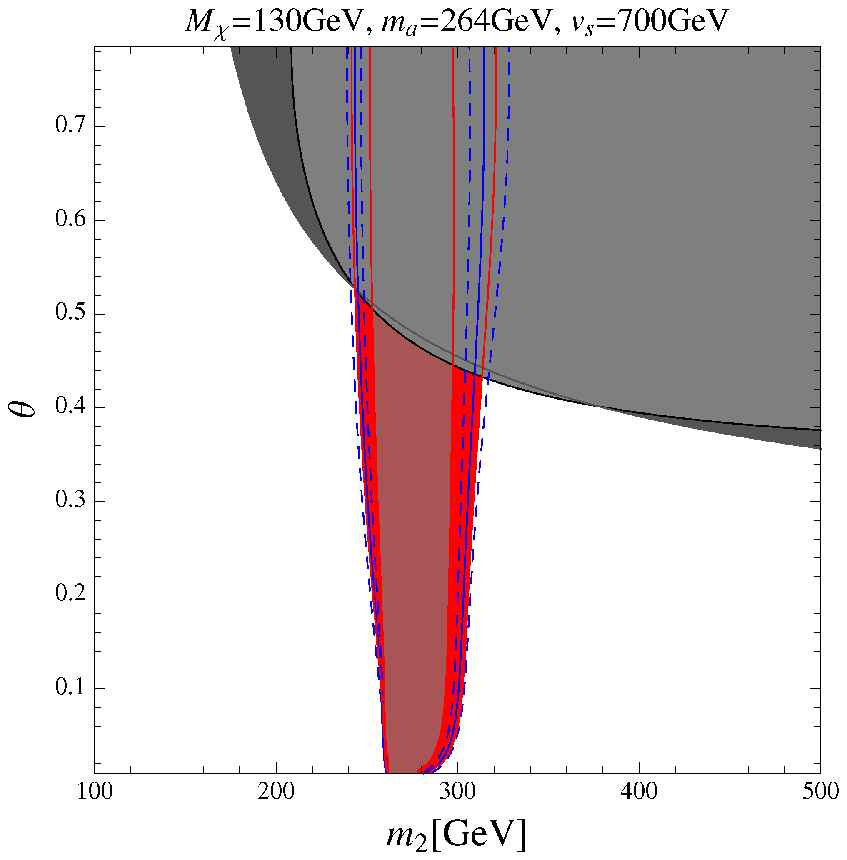}
\includegraphics[width=7cm]{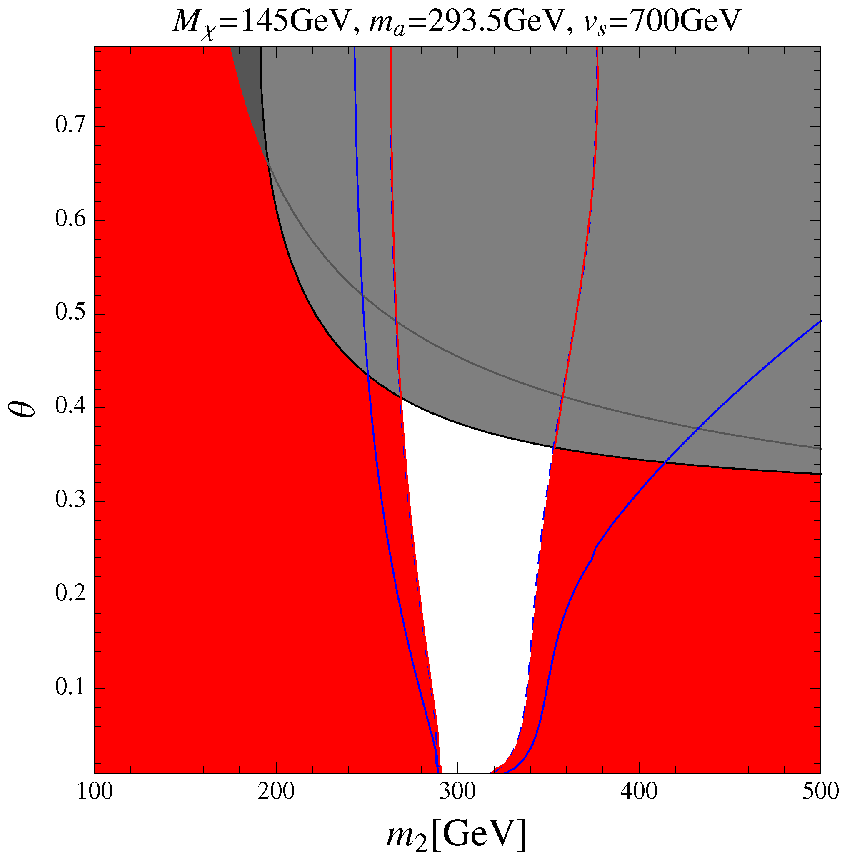}
\caption{Parameter space of the mixing angle of CP-even scalars vs the mass of singlet-like CP-even scalar.  The region consistent with
WMAP $3\sigma$ band of the relic density is bounded by blue dotted lines while the blue solid line is the central value within $3\sigma$. Left: Parameter space of the present partial annihilation cross section into a photon pair within $4-8\%$ (Red region bounded by red solid lines) and less than $4\%$ (Pink) and $c_1=c_2=1$. Right: Parameter space of the present partial annihilation cross section into $Z\gamma$ within $8-13\%$ (Red region bounded by red solid lines) and $c_1=0.2, c_2=1$. The mass of the Higgs-like scalar is chosen to $m_1=125\,{\rm GeV}$.
In both figures, gray region is excluded by XENON 100T and dark gray region is disfavored by the electroweak precision data. }
\label{fig:detection}
\end{figure}
%%%%%%%%%%%%% End   OF FIGURE ################

From the scattering process with the fermion dark matter, the spin-independent cross section for dark matter with nuclei is at tree level given by
\be
\sigma^{\rm SI}_{\chi-N}=|\lambda_\chi|^2\sin^2(2\theta)\,\cdot\frac{m^2_r f^2_N}{8\pi }\Big(\frac{m_N}{v}\Big)^2\Big(\frac{1}{m^2_1}-\frac{1}{m^2_2}\Big)^2
\ee
where $m_r\equiv m_N M_\chi/(m_N+M_\chi)$ is the reduced mass, $m_N$ is the nucleon mass, $f_N$, parametrizing the Higgs-nucleon coupling, is given by the sum of the light quarks ($f_L$) and heavy quarks ($f_H$) as $f_N=\sum f_L +3\times \frac{2}{27} f_H$ \cite{lebedev2}, 
and $m_{1,2}$ are physical CP-even scalar masses given in eq.~(\ref{masses}).
For instance, the direct detection bound is $\sigma^{\rm SI}_{\chi-N}\lesssim10^{-8}\,{\rm pb}$ for $M_\chi=130\,{\rm GeV}$ \cite{xenon}.  We note that the spin-dependent cross section with axion exchange is velocity-suppressed so the bounds coming from IceCube \cite{icecube} and Super-Kamiokande \cite{superK} does not constrain our model.

On the left plot of Fig.~\ref{fig:detection}, in the parameter space of the mixing angle and the mass of the singlet-like CP-even scalar, for $M_\chi=130\,{\rm GeV}$, we considered the WMAP $3\sigma$ band for the relic density and the branching fraction of the annihilation cross section into a photon pair less than $8\%$
and also show the limits from DM direct detection as well as the electroweak precision data (EWPD) at $95\%$ C.L. \cite{EWPD,lebedev}. We have also shown the case with $M_\chi=145\,{\rm GeV}$ on the right plot of Fig.~\ref{fig:detection} where both the WMAP bound and the gamma-ray line are obtained in all the parameter space away from the CP-even scalar resonance at $m_2\sim 2M_\chi$ and the direct detection bound is little changed as compared to the case $M_\chi=130\,{\rm GeV}$.
We can see that there is a parameter space with the mixing angle smaller than about $0.4-0.5$ that explains the Fermi gamma ray line at $E_\gamma=130\,{\rm GeV}$ and is compatible with all the phenomenological constraints.
As shown in both plots of Fig.~\ref{fig:detection}, the region below $m_2\sim 180\,{\rm GeV}$ is not constrained by direct detection because there is a cancellation effect in the scattering cross section due to the opposite signs in the amplitudes with CP-even scalars. 
In the range of the singlet-like CP-even scalar masses away from the cancellation zone, the mixing angle is constrained to be smaller than about $\theta=0.4$ by DM direct detection. We note that the EWPD does not give a stronger bound that the direct detection, in the region with the correct relic density and the Fermi gamma ray line. But, the former gives a stronger bound than the latter for $m_2\lesssim 260\,{\rm GeV}$ and $m_2\gtrsim 350\,{\rm GeV}$.

We note that if the singlet-like CP-even scalar gets mass outside the range, $122\,{\rm GeV}<m_2<128\,{\rm GeV}$, and there is no additional decay mode other than the ones of the SM Higgs boson, the LEP and LHC limits on the Higgs-like couplings will apply.
The LEP restricts the Higgs-like coupling to $Z$-boson,  $\xi^2\equiv \Big(g_{hZZ}/g^{\rm SM}_{hZZ}\Big)^2$ to be less than $0.5$ below $m_2<114\,{\rm GeV}$ \cite{lepHiggs}. So, the LEP limit constrains the low mass region for which there is no bound from the current direct detection. Furthermore, there is a strong limit from the 2011 year data at the LHC on the Higgs-like scalar below $200\,{\rm GeV}$, restricting the mixing angle to be as small as $\sin^2\theta=0.1$ \cite{lhcHiggs}.  
Thus, close to the cancellation zone, the LEP and LHC bounds on the mixing angle are much stronger than the direct detection bound.  On the other hand, away from the cancellation zone, the latter can be as strong as the former above $m_2=300 \,{\rm GeV}$. 

On the other hand, for $m_2>2m_1$ or $m_2>2M_\chi$, the singlet-like scalar can also decay into a pair of Higgs-like scalar or fermion dark matter with decay rates shown in appendix A. For a sizable mixing determined by the Higgs portal coupling $\lambda_{HS}$ and/or a sizable coupling between dark matter and the singlet scalar, the additional decay modes can easily dominate the SM Higgs decay modes of the singlet-like scalar: $\lambda_{HS}\gtrsim 0.1$ or $\lambda_\chi \gtrsim 0.1$ above the $WW$ threshold \cite{lebedev,lebedev2}. In particular, the dark matter coupling $\lambda_{\chi}$ to the mediator can be sizable unless heavy fermion mass running in loops is not too heavy. In this case, the current LHC limit would not be directly applied.

\section{A UV complete model}

The toy model given in Section 2 contains dimension-5 interaction terms between the axion-like scalar and the electroweak gauge bosons.
In this section, we consider the PQ symmetric extension of the MSSM\footnote{Different realizations of the NMSSM with PQ symmetry have been proposed recently \cite{pqnmssm} where SUSY breaking sector respects PQ symmetry. But, in our case, SUSY breaking sector breaks PQ symmetry while it respects discrete R symmetry. } as a UV completion of the toy model for generating the electroweak PQ anomalies. In this model, we can generate the electroweak PQ anomalies by the superparticles of the Higgs doublets, the Higgsinos, without introducing extra SM non-singlets. 
In this UV completion, there are the invisible axion multiplet $A$ and extra singlets, $S$ and $\chi_{1,2}$, as well as two more singlets, $X$ and $Y$, for the hidden sector SUSY breaking.

Assuming that the singlets have nonzero PQ charges
and R-charges as in Table 1, the superpotential for the extended Higgs sector is given by the following,
\be
W_{\rm vis}=\lambda_h S H_u H_d+ \lambda_\chi S \chi_1 \chi_2\,.
\label{superp}
\ee
The first term provides the $\mu$ term after the singlet $S$ gets a VEV while the second term leads to the axion coupling to a Dirac fermion dark matter composed of $\chi_1$ and $\chi_2$ for a nonzero singlet VEV. In our model, the QCD anomaly is generated only by heavy quarks coupled to the invisible axion multiplet $A$. PQ symmetry transformations are defined as $A\rightarrow A+i\theta_{\rm PQ}$ and $\Psi_i\rightarrow e^{iq_i \theta_{\rm PQ}}\Psi_i$ for all the other chiral superfields with PQ charge $q_i$ in the model.
The model is similar to KSVZ axion models, because there is no direct coupling of the invisible axion to the Higgs doublets.

\begin{table}[ht]
\centering
\begin{tabular}{|c||c|c|c|c|c|c|c|c|c|c|c|c|}
\hline 
& $Q, L$ & $\bar{U}, N$ & $\bar{D}, \bar{E}$ & $H_u$ & $H_d$ & $S$ & 
$\chi_1$ & $\chi_2$ & $X $ & Y \\ [0.5ex]
\hline 
PQ & $q_1$ & $0$ & $-q_1+q_2$ & $-q_1$ & $-q_2$ & $q_1+q_2$ & $q_\chi$ & $-q_\chi-q_1-q_2$  & $q_X$  & $2(q_1+q_2)$ \\ [0.5ex]
\hline
$Z^R_4$ & $1$ & $1$ & $1$ & 0 & $0$ & $2$ & $0$ & $0$  & $2$ & $0$ \\ [0.5ex]
\hline
\end{tabular}
\caption{PQ and $Z^R_4$ charges. Right-handed neutrinos $N$ are also included for neutrino masses.}
\label{table:U1charges}
\end{table}

Discrete R symmetries provide a solution to the $\mu$ problem and guarantees the proton stability in the MSSM \cite{Rsym1, Rsym2}. In particular, for $SU(5)$-GUT compatible R-charges, $Z^R_4$ and $Z^R_8$ are the only discrete R symmetries that allow for a singlet extension of the MSSM \cite{Rsym2}. A cubic interaction and/or linear term for the singlet $S$ is consistent with those discrete R symmetries but they are forbidden by PQ symmetry in our model.
A non-perturbative dynamics in hidden sector might generate the $\mu$ term and other PQ breaking singlet mass terms in the superpotential. But, in our model, we assume that the dominant PQ symmetry breaking in the Higgs sector comes from a spontaneous PQ breaking at high scale.

After the singlet $S$ gets a VEV, the singlet fermions, $\chi_1$ and $\chi_2$, obtain a Dirac mass as $M_\chi=\lambda_\chi \langle s\rangle$. If this singlet Dirac fermion is the lightest superparticle, it can be dark matter with odd R-parity as in the toy model in Section 2. Then, the Higgsinos generate the electroweak anomaly couplings to the axion coming from $S$ as follows,
\be
c_1=  c_2= -\frac{1}{2}(q_1+q_2).
\ee 

Now we discuss a microscopic model for SUSY breaking to generate the soft mass parameters.
Assuming that SUSY breaking multiplets $X,Y$ have PQ charges $q_X$ 
and $q_Y=2(q_1+q_2)$, respectively, the effective superpotential for the SUSY breaking sector is 
\be
W_{\rm hid}=\mu^2_1 X e^{-q_X A}+W_0 (1+\mu^2_2 Y e^{-q_Y A}).
\ee
Here, we have introduced the R-symmetry breaking\footnote{A possible domain-wall problem could arise after R-symmetry breaking of order $10^{11}\, {\rm GeV}$ related to $W_0$ in gravity mediation, but it depends on the reheating temperature after inflation. As far as the bound on the reheating temperature coming from gravitino problem, about $10^9 \,{\rm GeV}$, is satisfied, there is no domain wall produced after inflation, along the line of discussion in Ref. \cite{Rsym1,Rsym2}.}  in terms of a constant superpotential $W_0$.
After the saxion is stabilized, the above superpotential leads to nonzero F-terms, $F_X=\mu^2_1 e^{-q_X A}$ and $F_Y=W_0 \mu^2_2 e^{-q_Y A}$. Therefore, SUSY breaking sector also breaks PQ symmetry after the scalar partner of the axion multiplet is stabilized \cite{susybreakspq}.
Henceforth we take two F-terms to be comparable.
Furthermore, we can write the PQ and R invariant effective interactions composed of the invisible axion and SUSY breaking fields as follows,
\be
 \int d^4 \theta\, \frac{\alpha}{2M} Y^\dagger S^2+\int d^2\theta \,a X e^{-(q_X-q_1-q_2)A} H_u H_d+
\int d^4\theta \frac{b}{M}Y^{\dagger}X e^{-(q_X-q_1-q_2)A} S+{\rm h.c.} \label{hdo}
\ee
with $M$ being the messenger scale. 
The first term corresponds to a supersymmetric singlet mass term of order scalar soft mass from nonzero $F_Y$ term as follows,
\be
\Delta W_{\rm vis} = \frac{1}{2}\mu_S S^2, \quad\quad \mu_S=\alpha\, \frac{F^\dagger_Y}{M}.
\ee
This supersymmetric singlet mass is crucial to make the extra singlet fermion to be heavier than the Dirac fermion dark matter.
On the other hand, from eq.~(\ref{hdo}), the second term generates a B-term for Higgs doublets while the third term generates a singlet linear soft mass term. 
We note that the soft trilinear term is also generated by SUSY breaking with $\int d^2 \theta \frac{c}{M} Y\,e^{-q_Y A} SH_u H_d$ but they are suppressed as compared to soft scalar masses of order $\frac{|F_Y|}{M}\sim\frac{|F_X|}{M}$, because the SUSY breaking fields carry nonzero PQ charges. 
Nonetheless, gravity mediation can generate soft mass terms of order gravitino mass corresponding to all the terms present allowed in the superpotential. 
Therefore, we get the soft SUSY breaking terms as follows,
\bea
-{\cal L}_{\rm soft}&=&m^2_{H_u} |H_u|^2 + m^2_{H_d} |H_d|^2 +m^2_S |S|^2+m^2_{\chi_1}|\chi_1|^2+m^2_{\chi_2}|\chi_2|^2+\Big(\frac{1}{2}B_S \mu_S S^2+{\rm h.c.}\Big)
 \nonumber \\
&&+(\lambda_h A_\lambda S H_u H_d+\lambda_\chi A_\chi S\chi_1 \chi_2+B_\mu \mu H_u H_d+B_\chi \mu \chi_1\chi_2 +B_S m^2 S+{\rm h.c.}) \label{softmass}
\eea
where $A_\lambda$ is of order gravitino mass and 
\bea
B_\mu \mu \sim B_\chi \mu\sim a F_X e^{-(q_X-q_1-q_2)A},\quad  B_S m^2\sim b \frac{F^\dagger_Y F_X}{M}\,e^{-(q_X-q_1-q_2)A}
\eea
If the above B-terms are of order the soft scalar mass, for $|F_X|\sim |F_Y|$, we need
\be
a \sim  b\sim \frac{|F_X|}{M^2}.
\ee

PQ breaking linear soft mass for the singlet stabilizes the singlet at a nonzero VEV and gives the $S$ axion a nonzero mass.  
The minimization of the potential leads to \cite{nmssmreview}
\bea
\sin 2\beta&=&\frac{B_\mu \mu+(2A_\lambda+\lambda_h \mu_S)\langle s\rangle}{m^2_{H_u}+m^2_{H_d}+2\mu^2_{\rm eff}+\lambda^2_h v^2}, \quad \mu_{\rm eff}\equiv\lambda_h \langle s\rangle,\\
\langle s\rangle&=& \frac{\lambda_h v_u v_d (A_\lambda+2\mu_S)-B_S m^2}{m^2_S+B_S \mu_S+\mu^2_S+\lambda^2_h v^2}.
\eea
The scalar potential in NMSSM is more predictive because the Higgs quartic coupling is given by the gauge coupling and there is no singlet self-coupling.
However, as in the toy model, a mixing between CP-even Higgs and singlet is possible due to the singlet coupling to the Higgs doublets. 

The A-term in eq.~(\ref{softmass}) and the F-term for the $S$ singlet from the superpotential, $W_{\rm vis}+\Delta W_{\rm vis}$, can mix between the pseudo-scalar Higgs and the $S$ axion, so the tree-level DM annihilation into a pair of SM particles through the mixing could be large. Thus, in order to explain the gamma-ray line with the correct relic density, the mixing between the pseudo-scalar Higgs and the $S$ axion should be suppressed.
From the gamma-ray constraint that the extra tree-level axion mediation through the mixing is smaller than the one-loop induced counterpart, if the Higgs pseudo-scalar and $S$ axion masses, i.e. $m_A$ and $m_a$,  are comparable, we need to impose the soft mass parameter as 
\be
\frac{|A_\lambda-\mu_S|}{m_a}\lesssim \sqrt{\frac{g_2 M_\chi}{32\pi^2 v_s (m_a+M_\chi)}}\approx\sqrt{\frac{g_2 M_\chi}{96\pi^2 v_s } } \simeq 0.01.
\ee
If the pseudo-scalar Higgs mass is much larger than the $S$ axion mass, the amount of a tuning on $|A_\lambda-\mu_S|$ is reduced by a factor of $\frac{m_a}{m_A}$.
Then, the $S$ axion couples dominantly to the electroweak gauge bosons in the SM through the anomalies, playing a role of the mediator between dark matter and the SM.
Ignoring the mixing of pseudo-scalars for $A_\lambda \approx \mu_S$, we obtain the pseudo-scalar masses as \cite{nmssmreview}
\bea
m^2_A&=&\frac{2(B_\mu \mu+A_\lambda\langle s\rangle+\lambda_h \mu_S )}{\sin 2\beta}, \\
m^2_a&=&\frac{1}{\langle s\rangle}\Big(\lambda (A_\lambda+\mu_S)v_u v_d -B_S m^2\Big)-2B_S \mu_S.
\eea

We note that there are extra fields from the dark matter and messenger sectors in the supersymmetric models as compared to the toy model in Section 2: an extra Higgs doublet, the scalar partners of Dirac fermion dark matter and the fermionic partner of the $S$ singlet. 
But, the scalar partners of Dirac dark matter are not relevant for PQ and electroweak symmetry breaking. Furthermore, the extra singlet superparticles can be heavier than dark matter with mass $M_\chi$ so the additional annihilation channels into a singlet pair can be kinematically suppressed. Moreover, since the extra CP-even Higgs is heavier than the SM-like Higgs, the DM annihilation cross section is determined dominantly by the mediation channels with the $S$ axion and lighter CP-even scalars as discussed in our toy model in Section 2.

\section{Conclusion}

We have considered a Dirac singlet fermion as dark matter that communicates with the SM by a complex scalar mediator. Identifying a $U(1)$ global symmetry in the dark sector with PQ symmetry, a spontaneous breakdown of PQ symmetry at high scale generates a soft PQ-breaking mass of weak scale for the axion part of the complex scalar in a CP-invariant fashion. After the complex scalar gets a VEV, the effective axion interactions to the electroweak gauge bosons are generated by anomalies in the presence of extra heavy fermions with axion coupling.
This opens up a possibility that the axion mediates dark matter interactions close to the resonance, $m_a\sim 2M_\chi$, such that dark matter annihilates into $\gamma\gamma$ and $Z\gamma$ with sizable branching fractions while reproducing the relic density.
If the Fermi gamma ray line is confirmed, there will be interesting signatures to be pursued for DM direct detection and Higgs-like scalar searches below $300\,{\rm GeV}$ at the LHC in a near future. 

We also have presented a ultra-violet complete model that accommodates the axion coupling to the electroweak gauge bosons naturally by the singlet coupling to Higgsinos in the NMSSM with PQ symmetry. The SUSY extension relies on the specific PQ breaking soft mass terms in the NMSSM that are obtained in the presence of a discrete R symmetry. It would be worthwhile to investigate the implications of the gamma ray line on the Higgs boson and SUSY searches in this context.

\section*{Acknowledgments}
We would like to thank Marco Cirelli and Geraldine Servant for discussions. The implementation of our model with FeynRules became possible, thanks to the help from Neil  D. Christensen and Claude Duhr and through MC4BSM 2012 workshop. The work of HML and MHP is partially supported by a CERN-Korean fellowship. WIP is supported in part by Basic Science Research Program through the National Research Foundation of Korea(NRF) funded by the Ministry of Education, Science and Technology(2012-0003102).

\def\theequation{A.\arabic{equation}}

\setcounter{equation}{0}

\vskip0.8cm
\noindent
{\Large \bf Appendix A:  Decay rates}
\vskip0.4cm
\noindent
In the presence of the anomaly interactions between the axion and the electroweak gauge bosons, $c_{V_1 V_2}a\,\epsilon_{\mu\nu\rho\sigma} F^{\mu\nu}_{V_1}F^{\rho\sigma}_{V_2}$,  the decay rate for $a\rightarrow V_1 V_2$ with two gauge bosons, $V_1$ and $V_2$, having masses $M_1$ and $M_2$, respectively, is
\be
\Gamma(a\rightarrow V_1 V_2)=\frac{m^3_a}{2\pi}\,s_V |c_{V_1 V_2}|\Big(1-\frac{(M_1+M_2)^2}{m^2_a}\Big)^{3/2}\Big(1-\frac{(M_1-M_2)^2}{m^2_a}\Big)^{3/2}
\ee
with $s_V$ being the symmetry factor for the final states, for instance, $s_V=N!$ for $N$ identical final states. Furthermore, if the axion mass is larger than twice the dark matter mass, the axion can decay into a dark matter pair.
Then, the total decay rate of the axion is given by
\be
\Gamma_a=\Gamma_a(\gamma\gamma)+\Gamma_a(Z\gamma)+\Gamma_a(ZZ)+\Gamma_a(WW)+\Gamma_a({\bar\chi}\chi)
\ee
where
\bea
\Gamma_a(\gamma\gamma)&=&\frac{m^3_a}{\pi} |c_{\gamma\gamma}|^2, \\
\Gamma_a(Z\gamma)&=& \frac{m^3_a}{2\pi}|c_{Z \gamma }|^2\Big(1-\frac{M^2_Z}{m^2_a}\Big)^3, \\
\Gamma_a(ZZ)&=& \frac{m^3_a}{\pi} |c_{ZZ}|^2\Big(1-\frac{4M^2_Z}{m^2_a}\Big)^{3/2}, \\
\Gamma_a(WW)&=&  \frac{m^3_a}{2\pi}|c_{WW }|^2\Big(1-\frac{4M^2_W}{m^2_a}\Big)^{3/2}, \\
\Gamma_a({\bar\chi}\chi)&=&\frac{|\lambda_\chi|^2}{16\pi} \,m_a\Big(1-\frac{4m^2_\chi}{m^2_a}\Big)^{1/2}.
\eea
Here, the anomaly couplings are related to the original parameters in eq.~(\ref{action}) as 
\bea
c_{\gamma\gamma}&=& \frac{1}{16\pi  v_s}(c_1\alpha_1 \cos^2\theta_W+c_2\alpha_2 \sin^2\theta_W), \\
c_{Z\gamma}&=& \frac{1}{16\pi v_s} (c_2\alpha_2 -c_1\alpha_1)\sin(2\theta_W), \\
c_{ZZ}&=& \frac{1}{16\pi v_s}(c_2\alpha_2  \cos^2\theta_W+c_1 \alpha_1 \sin^2\theta_W),\\
c_{WW}&=& \frac{ c_2\alpha_2}{8\pi v_s}.
\eea

We also consider the decay rates of CP-even scalars. They can decay into the SM particles due to the mixing with Higgs boson as the SM Higgs does. Each partial decay width of the CP-even scalars into an SM particle pair is obtained from the one of the SM Higgs multiplied by $\sin^2\theta$ for $\tilde s$ and $\cos^2\theta$ for $\tilde h$. If kinematically allowed, the heavier CP-even scalar $\tilde s$ can decay into a pair of the lighter CP-even scalar $\tilde h$ and the CP-even scalars can decay into a dark matter pair by the direct coupling. 
So, the decay rates for the possible additional decay modes are
\bea
\Gamma({\tilde s}\rightarrow {\tilde h}{\tilde h})&=&\frac{\lambda^2_{HS}v^2}{8\pi m_2}\,\sqrt{1-\frac{4m^2_1}{m^2_2}},\\
\Gamma({\tilde s}\rightarrow {\bar\chi}\chi)&=& \frac{|\lambda_\chi|^2m_2 }{16\pi}\, \cos^2\theta \left(1-\frac{4m^2_\chi}{m^2_2}\right)^{3/2}, \\
\Gamma({\tilde h}\rightarrow {\bar\chi}\chi)&=& \frac{|\lambda_\chi|^2m_1}{16\pi}\, \sin^2\theta\left(1-\frac{4m^2_\chi}{m^2_1}\right)^{3/2}.
\eea

\def\theequation{B.\arabic{equation}}

\setcounter{equation}{0}

\vskip0.8cm
\noindent
{\Large \bf Appendix B:  Dark matter annihilation cross section}
\vskip0.4cm
\noindent

First, due to the dark matter coupling to the axion and the axion anomaly interactions,
the cross section times relative velocity for ${\bar\chi}\chi\rightarrow V_1 V_2$ is given by
\be
\sigma_{V_1V_2}v_{\rm rel}=\frac{1}{32\pi}\, s_V |\lambda_\chi|^2|c_{V_1 V_2}|^2\frac{s^2}{(s-m^2_a)^2+m^2_a \Gamma^2_a} \Big(1-\frac{(M_1+M_2)^2}{s}\Big)^{3/2}\Big(1-\frac{(M_1-M_2)^2}{s}\Big)^{3/2}
\ee
with $s$ being the center of momentum squared.
Then, the velocity averaged cross section for dark matter annihilation with axion mediation is
\be
\langle \sigma v\rangle_a = \langle \sigma v\rangle_{\gamma\gamma}+\langle \sigma v\rangle_{Z\gamma}+\langle \sigma v\rangle_{ZZ}+\langle \sigma v\rangle_{W W}
\ee
where 
\bea
 \langle \sigma v\rangle_{\gamma\gamma}&=&  \frac{1}{16\pi} |\lambda_\chi|^2|c_{\gamma\gamma}|^2\,\frac{16M^4_\chi}{(4M^2_\chi-m^2_a)^2+m^2_a\Gamma^2_a}, \\
  \langle \sigma v\rangle_{Z\gamma}&=&  \frac{1}{32\pi} |\lambda_\chi|^2|c_{Z\gamma}|^2\,\frac{16M^4_\chi}{(4M^2_\chi-m^2_a)^2+m^2_a\Gamma^2_a}\Big(1-\frac{M^2_Z}{4M^2_\chi}\Big)^3,  \\
  \langle \sigma v\rangle_{ZZ}&=&  \frac{1}{16\pi} |\lambda_\chi|^2|c_{Z Z}|^2\,\frac{16M^4_\chi}{(4M^2_\chi-m^2_a)^2+m^2_a\Gamma^2_a}\Big(1-\frac{M^2_Z}{M^2_\chi}\Big)^{3/2},\\
  \langle \sigma v\rangle_{W W}&=&\frac{1}{32\pi} |\lambda_\chi|^2|c_{WW}|^2\,\frac{16M^4_\chi}{(4M^2_\chi-m^2_a)^2+m^2_a\Gamma^2_a}\Big(1-\frac{M^2_W}{M^2_\chi}\Big)^{3/2}.
\eea

Second, dark matter can also annihilate into an SM particle pair by the Higgs portal interaction to the real scalar partner of the axion. The DM annihilations through the CP-even scalars are p-wave suppressed unlike the counterpart of axion mediation as shown below.
The dominant annihilation cross section coming from the CP-even scalar interaction is composed of
\be
\langle\sigma v\rangle_s=\langle\sigma v\rangle_{{\bar f}f}+\langle\sigma v\rangle_{WW}+\langle \sigma v\rangle_{ZZ}+\langle \sigma v\rangle_{{\tilde h}{\tilde h}}.
\ee
The partial annihilation cross section into an SM fermion pair is
\bea
\langle\sigma v\rangle_{{\bar f}f}&=&
\frac{3N_c |\lambda_\chi|^2 m^2_f \sin^2(2\theta)}{32\pi v^2}\,P_1 P_2
M^2_\chi
\left((m_1^2-m_2^2)^2+(m_1 \Gamma_1-m_2 \Gamma_2)^2\right)\nonumber \\
&&\quad\times\Big(1-\frac{m^2_f}{M^2_\chi}\Big)^{3/2}\frac{T}{M_\chi}
\eea
with $T$ being the temperature of the universe and $P_i\equiv [(4M^2_\chi-m^2_i)^2+m^2_i \Gamma^2_2]^{-1}$ (i=1, 2). 
From the gauge-Higgs interactions, $c_V h V_\mu V^\mu$ with 
\be
c_{W}=-2\frac{M^2_{W}}{v}\,,\quad c_{Z} = - \frac{M^2_{Z}}{v}\, ,
\ee
the annihilation cross section into a gauge boson pair is similarly obtained as follows, 
\bea
\langle\sigma v\rangle_{VV}&=&s_V\,\frac{3 |\lambda_\chi|^2 c^2_V\, \sin^2(2\theta)}{32\pi M^4_V}\, P_1 P_2M^4_\chi
\left((m_1^2-m_2^2)^2+(m_1 \Gamma_1-m_2 \Gamma_2)^2\right)\nonumber \\
&&\quad 
\times \Big(1-\frac{M^2_V}{M^2_\chi}+\frac{3}{4}\frac{M^4_V}{M^4_\chi}\Big)\Big(1-\frac{M^2_V}{M^2_\chi}\Big)^{1/2}\frac{T}{M_\chi}
\eea
with $s_V$ being the symmetry factor.
Finally, from the interactions between CP-even scalars, $\frac{1}{3!}a_1 \,{\tilde h}^3+\frac{1}{2}a_2\, {\tilde s}\, {\tilde h}^2$, the annihilation cross section into a Higgs-like pair is  
\be
\langle\sigma v\rangle_{{\tilde h}{\tilde h}}=
\frac{|\lambda_\chi|^2}{16\pi}\Big(1-\frac{m^2_1}{M^2_\chi}\Big)^{1/2}\frac{T}{M_\chi}\,
\Bigg(A_{tt}+A_{ts}+A_{ss} \Bigg) \, ,
\ee
where
\be
A_{tt}\equiv\frac{ |\lambda_\chi|^2   \sin^4{\theta} \, M_\chi^2  \left(9 M_\chi^4-8 M_\chi^2\, m_1^2+2 m_1^4\right)}{\left(2 M_\chi^2-m_1^2\right)^4}\, , \\[3mm]
\ee
\be
A_{ts}\equiv\frac{ \sin^2{\theta} \, M_\chi \left(5 M_\chi^2-2 m_1^2\right)}{\sqrt{2} \left(2 M_\chi^2-m_1^2\right)^2} 
\sum_{i=1,2}P_i\, {\rm Re}[\lambda_\chi  \tilde a_i^*] \left(4 M_\chi^2- m_i^2\right)
\, ,\\[3mm]
\ee
\be
A_{ss}\equiv \frac{3}{8}\left(\sum_{i=1,2}P_i |\tilde a_i|^2
+2P_1P_2\,{\rm Re}[\tilde a_1 \tilde a_2^*] \left\{(4M^2_\chi-m^2_1)(4M^2_\chi-m^2_2)+m_1 m_2 \Gamma_1 \Gamma_2 \right\}
\right)
\ee
with
\bea
&&\tilde a_1\equiv \frac{3 m_1^2}{v_s v}\left(v_s \cos^3{\theta}+v \sin^3{\theta}\right) \sin{\theta} \, , \\
&&\tilde a_2\equiv\frac{\sin\left(2\theta\right) \left(2m_1^2+m_2^2\right)}{2 v_s v} 
\left(v \sin{\theta}-v_s \cos{\theta}\right) \cos{\theta} \, .
\eea

\end{document}